\documentstyle [12pt] {article}

\begin{document}

 \bibliographystyle{unsrt}


\title{Comparison of
Spectral Method and Lattice Boltzmann Simulations of
Two-Dimensional Hydrodynamics}

\author{D.O.Mart\'{\i}nez, W.H.Matthaeus, S.Chen\thanks{
Center for Nonlinear Studies and Theoretical Division, MS-B213, Los Alamos
National Laboratory, Los Alamos, NM 87545},  and D.C.Montgomery\thanks{
Department of Physics and Astronomy, Dartmouth College, Hanover} \\
      Bartol Research Institute\\
University of Delaware, Newark, DE 19716
\\}

\maketitle
\date{  }

\begin{abstract}
We present numerical solutions of the two-dimensional Navier-Stokes equations
by two methods; spectral and the novel Lattice Boltzmann Equation (LBE) scheme.
Very good agreement is found for global quantities as well as
energy spectra. The LBE scheme is, indeed, providing reasonably
accurate solutions of the Navier-Stokes equations
with an isothermal equation of
state, in the nearly incompressible
limit.  Relaxation to a previously reported ``sinh-Poisson'' state is
also observed for both runs.
\end{abstract}
\vspace{0.3in}


\vspace{0.3in}

\pagebreak
\newpage

\section{Introduction}
In recent years\cite{fhp1,fhp2,wolf}, lattice gas models have
been developed for a number of fluid
and fluid-like systems. Interest in the study of these
methods derives from both theoretical and practical motivations.
On the one hand, lattice gases provide a novel perspective on
complex physical systems, differentiating the macroscopic physical
effects that are observable from the simplified microscopic properties
that ultimately are responsible for what is observed. From a
computational perspective, lattice gases have shown promise as an
alternative approach for solution of fluid-like partial differential
equations, especially on parallel computers where the relative independence of
the lattice nodes can be exploited for computational efficiency.
Lattice gas models of the Cellular Automaton (CA) type
have been developed for many systems, including hydrodynamics\cite{fhp1},
magnetohydrodynamics\cite{hudong-bill} (MHD), multi-phase flows\cite{roth1},
and flows through porous media\cite{roth2,schen2}.
Generally speaking CA models are plagued with noise, so that very large spatial
grids must be used. Schemes
to improve the situation are made difficult by complexity of collisions,
difficulty in eliminating spurious modes, lack of Galilean invariance,
and other problems.
Nevertheless, some appealing results have been obtained using CA fluid
models. Lattice Boltzmann (LBE) methods\cite{macn-zan,hig-jim}
represent an improvement in terms of noise and have produced
promising results in computations. Many of the previous demonstrations
of lattice gas have shown clearly that
complex physical fluid phenomena can be reproduced by these methods.
Examples include wave propagation in hydrodynamics and MHD,
vortex "streets" in viscous flows around obstacles,
immiscible fluid effects, and others.
However, most and perhaps all of these demonstrations have
been confined to a qualitative verification of the physics, and have stopped
short of showing that the lattice methods in fact provide an alternative
method for quantitatively accurate solution of the fluid equations.

Since the introduction of LBE methods there has been recognition
that the LBE framework provides opportunity to eliminate some, and perhaps
all of the fundamental problems in the lattice gas approach.
Two of these improvements are the use of the ``single time relaxation
approximation'' (STRA) collisions\cite{ccmm,qian}, and the ability to
introduce corrections to the pressure that modify the equation of
state and eliminate spurious modes\cite{qian,ccm}. In addition, the LBE
approach permits
greater flexibility in implementations in terms of lattice structure
and dynamical ``rules''.

Here we present a study of an improved LBE method in two dimensional
(2D) hydrodynamics,
demonstrating that for a simple nonlinear shear layer problem, the
LBE method is accurate and effective. Related study of the performance
of a 3D LBE scheme has been
presented by Chen et al\cite{schen4}.
The problem we choose to address is a simple 2D periodic shear layer,
perturbed by the addition of a low level of random ``noise''.
This basic flow problem is of broad relevance to flow applications
in geophysics, aerodynamics and space physics, and
 has been studied in laboratory situations and through numerical simulations.
Although three dimensional effects are absent in this treatment
of the nonlinear shear instability problem, and the 2D
physical phenomena are well known, the simplicity
and familiarity of this problem makes it a good
starting place for accurate demonstration of the LBE method.

The physical effects we are interested in reproducing are:
spectral transfer, energy and enstrophy decay, relaxation to the long time
``maximum entropy state'' at times shorter than viscous decay time.
Although our primary purpose is to examine the incompressible behavior,
the improved LBE scheme
also is seen to accurately provide information about the ``nearly
incompressible'' features of the dynamics, including waves and nearly
incompressible pressure and density fluctuations.

\section{Lattice Boltzmann Method}

We adopt a numerical scheme appropriate to 2D hydrodynamics
that is based upon the Lattice Boltzmann Equation, giving rise to
the abbreviation ``LBE'' method.
This approach to solution of fluid equations is based upon
the kinetic equations
associated with
Cellular Automaton (CA) models for
fluids[1,2,3]. In the CA formulation,
the Boltzmann equation does not enter into numerical implementations,
since it is constructed solely to demonstrate that certain averaged
functions of the lattice dynamics
approach solutions of the fluid equations. The LBE method arises
from the suggestion\cite{macn-zan} that a direct solution of these
equations would provide an alternative numerical approach to
computation, conceptually midway between the Boolean CA dynamics and
the continuum fluid equations.

The several types of LBE models proposed thus far share
with one another the advantage, relative to the underlying CA model, of
significantly reduced noise. However, it has also been recognized that
the LBE approach allows for simplifications to improve numerical efficiency,
as well as improvements to ``cure'' problems that arise in the underlying
CA models.
These refinements have substantially improved the prospect of
useful LBE computations.
The first major LBE refinement was the recognition that
the ``exact'' LBE collision integral is unnecessarily complex and numerically
inefficient\cite{hig-jim}.
The first idea for streamlining the evaluation of the collision integral
was\cite{hig-jim} to linearize the exact Boltzmann form.
Evidently such a simplification preserves the tendency to
approach the desired local equilibrium microscopic state, which is already
known (from CA theory) to lead macroscopically to hydrodynamics.
The only cost is a certain amount of departure of the distribution
from what it would be in the CA case. However, since the departures
from equilibrium are generally assumed to be small, this is not
expected to produce discrepancies in the physical results.
Expanding on the idea that the details of the collision operator need not
correspond to the Boltzmann approximation to the exact CA rules, two groups
nearly simultaneously offered the suggestion\cite{ccmm,qian,ccm}
that the exact collision operator can be, in effect, discarded,
provided that one adopts a collision operator that leads, in a
controllable fashion, to a desired local equilibrium state. By a
``desired'' equilibrium, we mean (1) one that depends only
upon the local fluid variables, which themselves can be computed
from the actual values of the local distribution at a point, (2) one that
leads to the desired macroscopic equations (e.g., the Navier Stokes equation),
and (3) one that admits whatever additional properties that are sought,
such as
simplicity or removal of nonphysical lattice effects.
Recent work has shown that property (2) can be maintained rather easily,
even when the collision operator departs significantly from the form taken in
the Boltzmann treatment of the CA. In fact such departures are desirable from
the point of view of several factors of type (3).

Chen et al\cite{ccmm} offered the first suggestion that one could use
the single time relaxation, or STRA,
collision operator
for a MHD LBE method.
Subsequently, a similar method\cite{qian} was described, and referred
to as a ``BGK'' collision integral, in reference to the
more elaborate collision treatment of Bhatnagar, Gross and Krook\cite{BGK}.
The essence of the suggestion for the LBE method is that
the collision term $\Omega(f)$ be
replaced
by the well known classical single time relaxation approximation
$$ \Omega(f) = -\frac{f-f^{eq}}{\tau}. \label{1}$$
An appropriately chosen equilibrium distribution is denoted by
$f^{eq}$, which depends
upon the local fluid variables, and a lattice relaxation time $\tau$ that
controls the rate of approach to this equilibrium.
Later, Qian et al\cite{qian} and Chen et al\cite{ccm} described a STRA
 method for hydrodynamics that incorporates
the form (1), but which also includes a reservoir of ``stopped'' particles
that enter into the equilibrium distribution to prevent the particle
distribution from ``cooling'' in regions of higher fluid speed. The latter
problem had plagued earlier CA implementations of fluid models by giving
rise to a velocity dependent pressure. An improper equation of state of this
kind introduces nonphysical
compressive effects \cite{DahlMontDool,schen5}, including
spurious oscillations, and incorrect pressure profiles in channel flows.
These effects are completely eliminated, to all orders in the
Mach number, by these ``pressure corrected'' LBE
schemes\cite{qian,ccm}. In contrast, multispeed CA models only
partially correct the equation of state, by moving the
velocity dependence of the pressure to higher order.
Still further improvements to the method were realized when the
stopped particle reservoir was parameterized in such a way\cite{accd}
that the sound speed could be controlled, enabling higher Mach number flows,
and
in principle, shocks, to be computed with the STRA-LBE scheme.

In the subsequent sections, we present results obtained with an LBE scheme for
2D hydrodynamics, that incorporates a number of the above described features.
We use a square lattice with eight moving particle directions
plus stopped ``particles''\cite{qian,koelman}.
In CA terminology, this ``9-bit model'' refers to a lattice
dynamical system in which particles stream from nodes on the lattice to
the nearest neighbor nodes at fixed speeds, experiencing collisions at
each node, which modify the particle state, and on average drive the particle
distribution toward equilibrium. Nearest neighbor nodes relative to a node
at $\bf x$ are located at the face-centers ${\bf x} + {\bf c}^I_a$,
for $a = 1,2,3,4$, with
${\bf c}^I_a\equiv (\cos{(a-1)\pi/2}, \sin{(a-1)\pi/2})$, and the
vertices of the square centered about $\bf x$, i.e.,
${\bf x} + {\bf c}^{II}_a$, for $a = 1,2,3,4$, with
${\bf c}^{II}_a\equiv \sqrt{2}(\cos{(a-1/2)\pi/2}, \sin{(a-1/2)\pi/2})$.
To move to the appropriate node during the streaming step,
a particle in state $I_a$ moves with velocity ${\bf c}^I_a$ while
particles in the state $II_a$ move with velocity ${\bf c}^{II}_a$.
In ``lattice units'', the lattice side can be taken to be $\delta x = 1$ and
the
lattice streaming time $\delta t = 1$, so that type $I$ particles have unit
speed and type $II$ particles have speed $\sqrt{2}$.

Turning to an LBE treatment of the dynamics, we
denote the moving particle distribution function by
$f^k_a$ for $k=I$ or $II$ and $a=1,2,3,4$, while the component of the
particle distribution referring to the stopped particles (which do not stream)
is designated by $f_0$.
Adopting a single time collision operator, along with the foregoing
streaming rules, we arrive at a kinetic equation
\begin{equation}
f_a^k({\bf x}+{\bf c}^k_a,T+1)-f_a^k({\bf x},T) =
 -\frac{f^k_a-f^{k (eq)}_a}{\tau}
\end{equation}
where $k=I$ or $II$.

Our LBE dynamical system is completed by choosing the
equilibrium distribution\cite{qian},
\begin{eqnarray}
f_0 &=& \frac{4}{9} \rho [1 - \frac{3}{2} u^2 ] \nonumber  \\
f_a^{I} &=& \frac{\rho}{9} [ 1 + 3 {\bf c}^I_a \cdot {\bf u} +
      \frac{9}{2} ({\bf c}^I_a \cdot {\bf u})^2 - \frac{3}{2} u^2 ] \\
f^{II}_a &=& \frac{\rho}{36} [ 1 + 3 {\bf c}^{II}_a \cdot {\bf u} +
      \frac{9}{2} ({\bf c}^{II}_a \cdot {\bf u})^2 - \frac{3}{2} u^2 ]
\nonumber
\end{eqnarray}
where the mass density $\rho$ and fluid velocity $\bf u$ are defined by
\begin{equation}
\rho = f_0+\sum_{k,a}f^k_a
\end{equation}
and
\begin{equation}
\rho {\bf u} = \sum_{k,a}{\bf c}^k_a f^k_a.
\end{equation}

Several fundamental properties of this LBE scheme can be
readily demonstrated, based upon the choice of equilibrium and the kinetic
equation (1). It is straightforward to show that
the pressure $p = C_s^2 \rho$, where $C_s = 1/\sqrt{3}$
is the sound speed. This is a ``pressure corrected'' LBE scheme with an exact
isothermal equation of state.
Next, considering moments of the kinetic equation, expanded according to a
multiple scale Chapman-Enskog procedure, we find that the long wavelength
low frequency behavior corresponds, at leading order, to a fluid equation
for the velocity
field.  In addition, if one invokes a low Mach number ordering, which allows
an approach to incompressible behavior, one obtains, in first approximation,
the incompressible
Navier Stokes equations,
\begin{eqnarray}
\frac{\partial {\bf v}}{\partial t} + {\bf v}\cdot {\bf \nabla v} =
- \frac{1}{\rho_0} {\bf \nabla} p^\infty+ \nu \nabla^2 {\bf v},
\end{eqnarray}
where $p^\infty$ is the incompressible pressure, and
$\rho_0$ is the conserved initial mean density.
Likewise, the small time dependent density variations obey a continuity
equation
\begin{equation}
\frac{\partial \rho}{\partial t} + {\bf \nabla} \cdot (\rho {\bf v}) = 0.
\end{equation}
At next order in the Chapman Enskog expansion,
making use of the STRA collision operator with constant relaxation
time scale $\tau$,
we find that the viscosity is
$\nu = (2 \tau -1)/6$  ($\tau > 0.5$),
which has the computationally
desirable property of being independent of the density.
Further discussion of the approach to incompressibility is given in section 6,
and some additional remarks concerning the viscosity and the
physical interpretation of $\tau$ are given in the Appendix.
\section{Shear Layer Simulations: Spectral and Lattice Boltzmann}
The idealized shear layer consists of uniform velocity reversing
sign in a very narrow region. That corresponds to a vorticity
$\omega = (\nabla \times {\bf v})_z $
different from zero only in the region of the sheared flow, and it is,
in the ideal situation, a delta function. Therefore, we generated our initial
conditions, in a simulation domain that is a square box of side
$2\pi$,  with a spectral representation of delta functions
at $y=\pi/2$ and $y=3\pi/2$ (with opposite signs) for the vorticity,
truncated to include the appropriate Fourier amplitudes
with wavevectors $k=1$ through $8$.
This configuration is steady in the absence of viscosity
and, although the simulations are viscous, we add
a perturbation to trigger the nonlinear terms of the
Navier Stokes equation. To this end the velocity Fourier modes
with $1\leq k \leq 60$ where excited with random phases and with
an energy spectrum of $k^{-3}$ for high $k$, and peaked at $k=3$.
This ``noise''
was such that added about $10\%$ of the energy already present
due to the idealized shear flow. In fact, a lower noise level
would be adequate
to trigger the nonlinear dynamics, but a larger level was used for
reasons that will be further explored in Section 5.
Thus, we have initially $E_k=0.5$,
$\Omega=5.738644$ (Enstrophy), $P=0.1866146E+04$ (Palinstrophy)
and $Q=0.2375568E+07$ (Q-enstrophy). Proceeding from this initial
condition, the subsequent dynamics gives rise
to a familiar set of phenomena associated with the two
dimensional shear layer,
including vortex layer breakup, vortex rollup and coalescence of like-signed
vortices. These will be described for the spectral and LBE runs in the next
section.

Operationally, the spectral run is a standard type, familiar in turbulence
computations, using an Orszag-Patterson implementation of
a Fourier Galerkin scheme.
We use a periodic box of side $2\pi$ and resolution
$256^2$, which allows to use a Reynolds number $R=10,000$, and still
resolve the dissipation wavenumber.
This method solves the equation for the vorticity $ \omega=({\bf \nabla}
 \times {\bf v})_z$,
\begin{equation}
\partial \omega/\partial t + {\bf u} \cdot \nabla \omega =
 \nu \nabla^2 \omega,
\end{equation}
 using a fast transform evaluation of the nonlinear
couplings, along with appropriate procedures for removal of aliasing errors.
The Reynolds number for the longest wavelength is $\sim 1/\nu$.
Time integration is a second order explicit scheme, using fixed time steps
of $\Delta t = 1/1024$.
The characteristic time scale for the motion of the large scale
eddies is estimated from the energy as $T_{SP} = L/\sqrt{2E}$, for a
characteristic length scale $L$.
In view of the slow decay of the energy, we estimate $T_{SP}$ using the
initial value of the energy, and the unit length associated with the longest
wavelength in the periodic box.
Thus, in simulation units of time, the large scale eddy turnover time
is $T_{SP}\approx 1$. The system is evolved up to  simulation time $t=119$.
The spectral simulation begins with the specified Fourier coefficients that
generate the initial data, and the complete set of vorticity Fourier
coefficients are stored at later times for subsequent comparison with the LBE
results.


For the LBE run, we obtain the initial fields from the relation
${\bf v} = {\bf \nabla}\psi \times {\bf z}$, where ${\bf z}$
is the unit vector in the $z$ direction, taking the stream function
$\psi$ from the solution to $\nabla^2 \psi = -\omega$, which is
algebraically solved in Fourier space using the appropriate value of
$\omega$ from the spectral run.
The initial density is
set to a constant. These fields are then used to initialize the
distribution function $f$ to its equilibrium value, for these specified
fields, using Eq.(2).
After this initialization procedure, the LBE system is evolved
by subjecting it to the sequence
of streaming and collisions alluded to above.

For the LBE simulation we used a $512^2$ box. For $2D$ hydrodynamics
turbulence an estimation of the dissipation length scale $L_d$ can be
obtained with $L_d/L_0 \sim R^{-1/2}$, where $L_0$ is the energy-containing
length. For $L_0=512$ we get $L_d \sim 5$, that is, the viscous dissipation
mechanisms are effectively occurring in a scale of the order of five cells.
A smaller run of $256^2$ size was carried out; however effects
typical of lack of resolution of the dissipation length were observed for this
computation. Having chosen the appropriate lattice size
and $\sqrt{\langle u^2\rangle}=0.04$
for the LBE system, the relaxation parameter $\tau$ is
then fixed to give the proper viscosity value to  obtain $R=10,000$, using
the expression $\nu=(2\tau -1)/6$ (in lattice units).
In particular
to achieve an LBE Reynolds number, i.e $R = U L / \nu = 10,000$,
we use
\begin{eqnarray}
R = 0.04 \times \frac{512}{2 \pi} \times\frac{1}{(2\tau-1)/6} \\ \nonumber
\end{eqnarray}
arriving as $R=10,000$ when $\tau=0.500977848$.
Although the physical Reynolds numbers will be time dependent,
scaling with the characteristic fluctuating fluid velocities, we expect that
these LBE parameters produce Reynolds numbers in the two types of runs
that are within 10\% in value.

To be able to compare LBE and spectral runs we also have to relate
the time units, from lattice convection
units (i.e., time needed to propagate microscopic information
from cell to cell) to large scale eddy turnover time. That conversion is
done in the following way.
Using characteristic lengths and velocities for both spectral and
LBE systems, we can find a relationship between the characteristic
times for the schemes. Thus, using $L_{LBE}=512$, $L_{SP}=2\pi$,
$U_{LBE}=\sqrt{\langle u^2\rangle}=0.04$, $U_{SP}=1$, we can get an expression
connecting the typical time for evolution of both systems,
\begin{equation}
\frac{T_{LBE}}{T_{SP}} = \frac{L_{LBE}/L_{SP}}{U_{LBE}/U_{SP}}=
\frac{512}{2 \pi} \frac{1}{0.04},
\end{equation}
thus
\begin{equation}
T_{LBE} = 2037.12 \; \;T_{SP}\nonumber
\end{equation}
so we need about $2037$ LBE time steps to complete one spectral
characteristic time.

Global features of the evolution are calculated dynamically
for the LBE computation every $200$ LBE-time steps (i.e. about $1/10$
eddy turnover time). The velocity field is scaled to the units
used for the spectral simulation and is Fourier transformed. The
vorticity $\hat\omega(k)=i({\bf k \times \hat v(k)})_z$ is then evaluated,
from which the energy, enstrophy, palinstrophy, q-enstrophy and mean
square stream function are computed. This scheme assumes that
the fluid is incompressible, i.e., for  these global diagnostics, and for
energy wavenumber spectra, the (small) admixture of nonvortical
velocity fluctuations associated with the compressible LBE dynamics, is
ignored.
In Section 6 we will further discuss
the validity of this approximation.

\section{Comparison of Spectral and LBE results}
In spite of the organized large scale appearance our initial data -
a periodic shear layer perturbed by random fluctuations, the evolution of
the system in time is quite typical of 2D incompressible hydrodynamics.
Thus, the time histories of global quantities, illustrated in Fig. 1, are
familiar in appearance and interpretation. Fig. 1 shows the evolution of the
energy $E= \sum_{\bf k} |\omega ({\bf k})|^2/k^2$, the enstrophy
$\Omega = \sum_{\bf k} |\omega ({\bf k})|^2$, the Palinstrophy
$P = \sum_{\bf k} k^2|\omega ({\bf k})|^2$, and (lacking a better nomenclature)
the ``Q-enstrophy'' $Q = \sum_{\bf k} k^4|\omega ({\bf k})|^2$.
(The sums are over the independent wavevectors ${\bf k}$.)

$E$ and $\Omega$ are inviscid invariants, and therefore are monotonically
decreasing in these dissipative simulations. On the other hand $P$ and $Q$
can be amplified as well as dissipated, and are not monotonic.
One can also prove that $\Omega/E$ decreases in time\cite{ting}.
This is associated with the tendency for 2D Navier Stokes flow to
engage in ``selective decay'', wherein the turbulence drives the spectrum
towards its geometrically determined extremal state $\Omega = K_{min}^2
E$, where $K_{min}$ is the lowest allowed value of wavenumber, in a time
short compared with the decay of the flow due to viscosity.
The perspective provided by Fig. 1 is consistent with prior results in showing
that selective decay is at least a qualitatively useful picture of 2D
turbulence. For example, focusing on Fig. 1a) and
1b), one sees that $E$ decays quite slowly compared with $\Omega$, allowing the
conclusion to be drawn that the energy is ``back-transferred'' in $k$, where
dissipation is slow. On the other hand, the flow tends to produce additional
amounts of $P$ (see Fig. 1c), an effect that accelerates the dissipation
of $\Omega$, since $\dot \Omega = -2 \nu P$.
In addition $Q$ is also amplified early in the run (Fig. 1d), and
is dissipated at later times along with $E$, $\Omega$ and $P$.
This complex process of spectral transfer, involving both direct transfer
to higher $k$, and backtransfer to lower $k$ is familiar in 2D flows
\cite{KraichnanandMont}, and is a consequence of a very large number of
nonlinear couplings each involving triads of wave vectors. These couplings,
as well as their symmetries that give rise to inviscid conservation of $E$ and
$\Omega$, are accurately simulated by the spectral method simulation
technique.
What is new in the panels of Fig. 1 is evidence that the LBE method tracks
the spectral method closely with respect to evolution of $E$, $\Omega$, $P$
and $Q$. Therefore, even though the LBE method does not involve
 a wavevector representation, or even the vorticity, in any direct way,
it evidently provides a representation of the Navier Stokes
dynamics that is accurate enough to preserve the subtleties of 2D nonlinear
spectral transfer.

Each of the
quantities $E$, $\Omega$, $P$ and $Q$,
provides a measure of the distribution of vorticity
over wavenumber, and those with higher powers of $k$ weight the short
wavelengths more heavily.
A careful inspection of Fig. 1a-d shows that the quantities that emphasize
the lower $k$ part of the spectrum are most similar in the LBE and spectral
runs. Evidently, the departures of the LBE from the incompressible spectral
method are greatest at the higher wavenumbers.
Nevertheless, even fine features of the spectral method evolution
of $P$ and $Q$ are also seen in the LBE curves for the same quantities.

Wavenumber spectra of the energy are compared in Fig. 2, for the spectral
and LBE results, at times $t = 0, 5, 49$ and $80$ (in simulation time units,
i.e., eddy turnover times computed in terms of the initial data).
Fig. 2a) shows, for the two runs, the initial spectra, which are
identical by construction. Local peaks at the lower wavenumbers are associated
with the initial shear layers, while the higher $k$ powerlaw is
due to the ``noise'' perturbation.
By $t=5$, a substantial amount of spectral evolution has occurred in both
LBE and spectral runs, but, as is shown in Fig. 2b), the energy
spectra for the two cases have remained
extremely close. The most noticeable departures are at the highest values of
$k$, as expected from the discussion in the previous paragraphs.
Very similar energy spectra are also seen at much later time, as is
illustrated in Fig. 2c) and 2d) at times $t=49$ and $t=80$.
In these latter two comparison plots one can see clearly that significant
amounts of back transferred energy persists in the longer scales at these late
times, and that this effect is accurately portrayed in the LBE run.

Perhaps the most striking verification of the accuracy of the
LBE run is found in the direct comparison of contour plots of the
LBE vorticity with the spectral method vorticity at the same physical times.
In Fig. 3 we show pairs of vorticity contour plots at four times.
While the times are given in simulation times, it should be noted that the
equivalent LBE time was computed from the calibration discussed in the previous
section. The early time state, at $t=1$,
is seen in Fig. 3a), which shows, in both the
spectral and LBE cases, that the initial shear layers have begun the familiar
process of vortex roll-up. The vorticity distribution is extremely similar in
 the
two cases.  By a later time ($t=5$, in Fig. 3b) the roll-up has progressed and
has produced individual vorticity concentrations. These subsequently convect
in the flow due to all the vortices, and mergers occur between
like-signed vortices due to ``vortex collisions''.
Once again, the plots from
the two methods show great similarity, even down to detailed structures
near regions of like signed vortex interactions. Note that the same values of
vorticity contours are used in performing the comparisons.
A distinctive vortex collision is captured by both methods at time $t=17$,
shown in Fig. 3c.

At later times, all the positive and negative vortex concentrations have
separately merged into a single pair of vortices. Fig. 3d shows this state,
computed in both the spectral and LBE runs, at $t=80$.
It is clear that the LBE method has succeeded in reproducing many of the
important dynamical features obtained by the spectral method, which
has been the standard method for turbulence.
These features include the evolution of bulk quantities, the form and evolution
of the wavenumber spectra, and the detailed features of vorticity contours,
including vortex rollup and subsequent mergers of like signed vortices.
We now turn to some more subtle features of the flow, which appear also
to be well represented by the LBE method.

\section{Relaxation to ``sinh-Poisson'' most probable state}
		An interesting by-product of the decaying turbulence
computation just described concerns the extent to which the two-vortex
quasi-steady final state has vortex shapes which coincide with those recently
seen at the end of a study of decaying two-dimensional turbulence reported
elsewhere.  A slight digression is required before it is possible to display
the relaxation products of the turbulent computation in a way that will make
this connection clear.
		It has long been realized that in decaying 2D
Navier-Stokes flow, enstrophy or mean-square vorticity decayed
rapidly compared to energy or mean-square velocity, for reasons that are well
known. The separation of the time scales increases with Reynolds number, and
had led to a conjecture that the relaxed state of decaying 2D
Navier Stokes turbulence
would be one in which the enstrophy was minimized relative to the remaining
energy\cite{ting,bre-hai,mon-tur-va,whm-montg}.
In such a state, the only excitations left in the energy
spectrum would be those in the longest wavelengths allowed by the boundary
conditions.  Qualitatively, such a ``selectively decayed'' state would
resemble, for example, the states shown in the two panels of Fig. 3d.
		Some time ago\cite{us1,us2}, a highly-resolved (512x512), high
Reynolds number (14,286, based on the largest allowed wavelength), and
long-time (about 400 large-scale eddy turnover times) 2D Navier
Stokes spectral-method
computation was carried out, in an effort to test the above-described
"selective decay" hypothesis.  In broad outline, the tests confirmed the
hypothesis, but examined closely, departed from it.  In particular, a scatter
plot of computed pointwise vorticity vs. stream function revealed not a
linear proportionality between the two, as the selective decay hypothesis
would suggest, but rather a hyperbolic-sinusoidal one, in which the observed
connection\cite{us3} was that the late-time vorticity and stream function were
related by
\begin{equation}
c \omega = \sinh(|\beta|\psi)
\end{equation}
where c and $\beta$ are constants.  The result was surprising, to the extent
that it had been predicted two decades ago\cite{joy-mon,mon-joy}
from a mean-field theory of
most probable states, not for a viscous Navier-Stokes continuum, but rather
for a large number of ideal, discrete, parallel line vortices.  The subject
had developed, with  both analytical and numerical solutions of the
``sinh-Poisson'' partial differential equation\cite{joy-mon,mon-joy}
that had been derived to
describe the most-probable, or maximum-entropy, states, and a good
bibliography is given by Smith\cite{smith}.
A reformulation of the maximum-entropy
theory had been given in the context of
magnetohydrodynamics\cite{mon-tur-va2,ambr-va}, and a
further development of the foundation of the Navier-Stokes basis for it will
be given elsewhere\cite{mon-shan-whm}.
		Our intent in this Section is simply to point out that even
this somewhat unexpected and perhaps exotic hyperbolic-sine connection
between stream function and vorticity has been reproduced accurately in the
present LBE computation.  In Fig. 4, we graph two correlation functions vs.
time, with the broken line referring to the spectral method computation and
the solid line to the LBE computation.  Shown in Fig. 4 are correlations
between vorticity and stream function (lower curves) and between vorticity
and the hyperbolic sine of $\beta$ times the stream function (upper curves),
where the constant $\beta$ is determined from a least-squares fit to the
computed data.  For any two functions f(x,y) and g(x,y), the correlation
C(f,g) is defined by
\begin{equation}
C(f,g) \equiv \frac{\langle (f-\langle f\rangle )(g-\langle g\rangle )\rangle}
{[\langle(f-\langle f\rangle)^2\rangle\langle(g-\langle
g\rangle)^2\rangle]^{1/2}}
\end{equation}
where the angle brackets $\langle\rangle$ denote a spatial average
over the entire box.
Thus for any two functions which are proportional, C will be equal to unity.
The approach to the "sinh-Poisson" prediction is seen not only to be far
superior for the computed data, but it will also be noticed that the LBE and
spectral method computations again track each other to a remarkable extent.
		We remark also that there is a problem, as yet unsolved, of
extracting the observed statistical mechanical distribution of the LBE
variables for a vortex distribution directly from the LBE dynamics, without
the necessity of detouring  through the Navier-Stokes approximation.  This
must at present stand as a challenge for theory; a solution would be highly
desirable as a logical link between the microscopic and macroscopic dynamics.

\section{Nearly incompressible hydrodynamics in the LBE scheme}
In the previous sections, evidence was presented that the LBE method
reproduces many of the essential dynamical features of the {\it incompressible}
Navier Stokes equations, as computed by a spectral method code based upon the
vorticity equation. In particular, we found that the solutions appear to be
quantitatively similar to one another.
What differences there are between the spectral and LBE results appear to be
most pronounced at the higher wavenumbers.
It is tempting to assign these discrepancies to ``error'' in the LBE
formulation, and conclude that the methods correspond well,
for most of the diagnostics of interest, at times of up to
tens, or perhaps a hundred or more, characteristic nonlinear times.

However, there remains the possibility
that the LBE scheme, is, in some sense, more accurate than this would suggest.
We refer here to the possibility that the LBE scheme, in effect is solving
a {\it compressible} set of fluid equations, and therefore, would be
expected to approximate solutions of the incompressible equations only in
an appropriately defined limiting sense.
In fact, the compressible Navier Stokes equation itself also possesses this
property. For suitably chosen initial data, and for small Mach numbers,
the solutions of the compressible fluid equations are expected
\cite{klainerman-majda} to approximate the solutions to the incompressible
equations for at least some finite time interval. Simulations
\cite{gosh-whm} of the
compressible equations of 2D hydrodynamics (with a polytropic equation
of state) have also led to
the suggestion that finite Reynolds number extends the realm of this
expectation, so that in some cases the ``nearly incompressible'' nature
of a decaying flow may persist permanently.
Since the LBE method is intrinsically compressible, we can reasonably expect
that it, too, will admit a range of parameters and time in which its solutions
approach the desired incompressible solutions. This, indeed, is what we have
seen in the previous section. However, there is also the prospect that some of
the
{\it departures} of the LBE solutions from the spectral method
incompressible solutions might be attributable to
the slight effects of compressibility. In that case, at least some fraction
of the differences
between the spectral and LBE solutions might not be
errors of numerical origin, but rather physical
effects that lie
outside the realm of the incompressible equations.
We briefly explore this possibility here, by examining whether the LBE results
are consistent with the expectations of nearly incompressible fluid
theory\cite{klainerman-majda}.

Equation (5), the incompressible equation for the velocity field,
assumes that the leading order velocity field, say, ${\bf v}_0$,
is divergenceless, $\nabla \cdot {\bf v}_0 = 0$ and the leading order
density is constant, say $\rho = \rho_0 = {\rm const.}$. Again assuming that
this limit to incompressibility is obtained, we find that the incompressible
pressure $p^\infty$ appearing in (5), must satisfy
\begin{equation}
\nabla^2 p^\infty = -\rho_0 \nabla \cdot ({\bf v}_0 \cdot \nabla {\bf v}_0)
\end{equation}
which is a consequence of the time independence of
$\nabla \cdot {\bf v}_0 = 0$.
On the other hand, prior to the limit to incompressibility, the
LBE system is found to obey the compressible equations
\begin{equation}
\rho \lbrack
\frac{\partial {\bf v}}{\partial t} + {\bf v}\cdot {\bf \nabla v} \rbrack =
 -{\bf \nabla} p + {\bf D},
\end{equation}
which, along with the continuity equation (6) and the equation
of state $p=C_s^2 \rho$ completes the specification of the long wavelength,
low frequency LBE dynamics.
The term {\bf D} on the right side of Eq. (14)
represents the viscous dissipation
terms for the {\it compressible} fluid limit of the LBE method. Since
neither the form nor the effects of
dissipative terms are central to the description of
near-incompressibility that we examine here, we neglect ${\bf D}$ in
the following discussion.

Let us define the
Mach number of the flow as $M \equiv \delta v/C_s$, with
$\delta v $ the rms value of ${\bf v}$. When $M << 1$ we expect there to be
conditions in which a decaying flow will remain nearly incompressible.
Klainerman and Majda\cite{klainerman-majda} have shown that the
additional required conditions are that the initial data satisfy
$\langle |\nabla \cdot {\bf v}|^2\rangle ^{1/2} = O(M)$ and $\delta \rho =
\langle(\rho - \rho_0)^2\rangle^{1/2} = O(M^2)$
where $\langle ...\rangle$ denotes a volume average.
In the LBE run discussed above, the initial $\delta v = 0.04$, and
$C_s = 1/\sqrt{3}$, so the initial $M=0.069$.  In addition, $\rho = \rho_0$
is uniform in the initial data. As for the velocity field, it is
computed for the LBE initially in terms of the real space values obtained from
the spectral method initial data. Thus, except for possibly errors due to
the finite LBE grid, it satisfies
$\nabla \cdot {\bf v} = 0$ initially.
Consequently, the conditions for the approach of the compressible equations
to the solutions of the incompressible equations appear to be well fulfilled.

In this circumstance, we expect that, for a finite time, the density should
remain ordered as $\rho = \rho_0 + M^2 (\rho^\infty + \rho^\prime) + O(M^3)$,
while the pressure (in convection speed units) should satisfy
$p = M^{-2}(p_0 + M^2(p^\infty + p^\prime) + O(M^3))$. Here, $p^\infty$
is the incompressible pressure, satisfying the Poisson equation (13).
There is also an additional pressure fluctuation
$p^\prime$, at the same order as the
incompressible pressure, but associated with acoustic waves, and decoupled
from the incompressible equation of motion.
The leading order density fluctuation is $\delta \rho \approx M^2(\rho^\infty
+ \rho^\prime)= \delta \rho^\infty + \delta \rho^\prime$,
where $\rho^\prime$ is also associated with acoustic waves.
In addition to the Poisson equation, the incompressible pressure
satisfies the relation $p^\infty + p^\prime \approx C_s^2 \delta \rho$.
In order for the incompressible dynamical equation to lack acoustic time scale
variations, we must apportion the leading order density fluctuations so that
$p^\infty = C_s^2 \delta \rho^\infty = M^{-2} \delta \rho^\infty$, the latter
equality making use of the convection speed units. Considering also the
velocity field, we note that, in a Fourier decomposition, we can readily
divide the velocity field as ${\bf v} = {\bf v}_L + {\bf v}_\perp$
where the longitudinal velocity ${\bf v}_L$ has $ \nabla \cdot {\bf v}_L \neq
0$ but $\nabla \times {\bf v}_L = 0$, while the transverse velocity
${\bf v}_\perp$ satisfies $ \nabla \cdot {\bf v}_\perp =0$
but $\nabla \times {\bf v}_\perp \neq 0$.
Then, for maintaining near-incompressibility we require
\cite{klainerman-majda}
that the solutions remain ordered so that ${\bf v}_L = O(M)$ for
(incompressible) convection speed units in which ${\bf v}_\perp = O(1)$.

The degree to which these expectations of nearly incompressible fluid theory
are seen in the LBE solution can be examined by analysis
of the LBE velocity and density fields. The results
are illustrated in the panels of Fig.\ 5.
The LBE velocity field is Fourier transformed and
decomposed into transverse
and longitudinal components by projections relative to wave vector $\bf k$.
Then, the rms transverse
velocity $U_\perp = \sqrt{\langle |{\bf v}_\perp|^2\rangle}$
and the rms longitudinal velocity
$U_L = \sqrt{\langle |{\bf v}_L|^2\rangle}$ are computed.
The relative magnitudes of $U_\perp$ and $U_L$ during the simulation
are shown in Figs.\ 5a and 5b. In Fig.\ 5a, the time history of $U_\perp/C_s$
is shown.  The value decreases slightly from its initial value, which is
very close to the value of the Mach number $M=0.069$ computed from the entire
velocity field. Thus, we expect that the longitudinal velocity is small, and
this is verified in Fig.\ 5b, which shows $U_L/U_\perp$ as a function of time.
The latter ratio meanders about a value near $0.010$.
Consequently, in convection speed units, it is clear that the condition
$U_L = O(M)$, required for the approach to incompressibility, is well
satisfied.

The density fluctuations may be decomposed as well,
in accordance with nearly
incompressible theory.
First, we simply evaluate the total density fluctuation $\delta \rho$, and
compare its value to the mean density and the Mach number
as time progresses.
This is shown in Fig.\ 5c as the solid trace, $\delta \rho/\rho_0$ vs. time.
We see that the magnitude of the root mean square
total density fluctuation is comparable to the
expected value, of order $M^2 \approx 0.0048$.
Next, we decompose the density into the part associated with the
underlying incompressible flow, and the part associated with acoustic activity.
Using only the transverse velocity field, we numerically solve
Eq. (13) for the incompressible pressure $p^\infty$.
The incompressible density fluctuation is computed
as $\delta \rho^\infty = M^2 p^\infty$. The root mean square
value of $\delta \rho^\infty$ is plotted also in Fig.\ 5c,
normalized to the mean density.
Again the result is clearly $O(M^2)$.
Finally, we compute the density fluctuation associated with
leading order acoustic effects through $\delta \rho^\prime =
\delta \rho - \delta \rho^\infty$ at each point in space.
Computing the root mean square $\delta \rho^\prime$ provides
a measure of the degree of acoustic activity. This
is illustrated as well in Fig.\ 5c, showing that this component of the
density fluctuation also remains of $O(M^2)$, again in accordance with
the expectations of nearly incompressible theory.

\section{Discussion and Conclusions}
In the above sections we have presented a detailed comparison of
solutions to the two dimensional Navier Stokes
equations obtained from a Lattice Boltzmann method and from a more
traditional spectral method.
The flow problem considered was a familiar shear layer initial value problem,
in periodic boundaries and prepared initially with a low level of random noise.
We find that the LBE method has provided a solution that is
``accurate'' in the sense that time histories of global quantities,
wavenumber spectra, and vorticity contour plots, are very closely
similar to those obtained from the spectral method. While the comparison
is best at early times, the solutions remain extremely close to one another
for at least several eddy turnover times, and in some ways remain
close for times up to a hundred turnover times.
In particular, details of the wavenumber spectra at high wavenumbers are
reproduced, as well as the detailed structure of vortex distributions seen
in the contour plots.
In addition, the LBE scheme faithfully reproduces the recently reported
long time
tendency for the stream function to approach a ``sinh-Poisson'' state
that emerges from a maximum entropy argument.
We have also explored the possibility that the LBE solution, to the
extent that it departs from a pure solution of the incompressible
equations, is remaining in the mathematically delineated regime
of ``nearly incompressible flow''. This indeed appears to be the case,
although a more complete verification would require comparison
with a fully compressible spectral algorithm, a refinement we have not as yet
undertaken.

It remains to discuss the accuracy of the LBE scheme in a quantitative way.
To do so we have computed several kinds of
normalized differences between the results of the two runs, which
are interpreted (for the most part) as errors in the LBE method.
The normalized errors in the bulk quantities,
energy, mean square stream function and enstrophy,
are computed, for example, as $|E_{SP}-E_{LBE}|/E_{SP}$, and shown in Table I.
(The suffixes $SP$ and $LBE$ refer to $\phi$ computed from the spectral
or LBE schemes, respectively.)
The normalized total rms error, defined for the spatially
dependent variable $\phi$ as
\begin{equation}
\varepsilon(\phi) = \left ( \frac { \langle (\phi_{SP} - \phi_{LBE})^2 \rangle}
                    { \langle \phi^2_{SP} \rangle } \right )^{\frac {1}{2}}
\end{equation}
This rms normalized error has been computed
as a function of time for
 $\phi$ taken as $\omega$, ${\bf v}_\perp$ or $\psi$.
In addition we have computed the kurtosis $K(\phi) =
\langle\phi^4\rangle/\langle\phi^2\rangle^2$
for $\phi$ taken as $\omega$, ${\bf v}_\perp$ or $\psi$.
Error in the kurtosis is conveniently expressed as $\Delta K/K =
|K_{SP}-K_{LBE}|/K_{SP}$, where the suffixes have the same meaning as above.
In table I we give the values of these normalized errors at spectral
method times $t = 1$, $10$, $50$ and $100$.

\vspace{7mm}
\begin{center}
\begin{tabular}{|c|c|c|c|c|c|}    \hline\hline
  &         &
\multicolumn{4}{c|}{TIME}  \\ \hline
Error             &          & $1$       & $10$      & $50$      &
$100$    \\ \hline
                  & $\psi$   & $0.00742$ & $0.04867$ & $0.35675$ & $1.12567$ \\
$\varepsilon$     & $v$      & $0.02136$ & $0.13685$ & $0.38283$ &
$1.21901$ \\
                  & $\omega$ & $0.12751$ & $0.53799$ & $0.65278$ & $1.37693$ \\
\hline\hline
                  & $\psi$   & $0.00097$ & $0.00623$ & $0.00984$ & $0.00161$ \\
$\Delta K/K$      & $v$      & $0.00297$ & $0.01966$ & $0.03168$ &
$0.08017$ \\
                  & $\omega$ & $0.00237$ & $0.01245$ & $0.05960$ & $0.05869$ \\
\hline\hline
                  &$\langle \psi^2 \rangle$
                             & $0.00043$ & $0.00957$ & $0.01676$ &
$0.01843$ \\
$\Delta \Phi/\Phi$& $E$      & $0.00081$ & $0.00075$ & $0.00045$ & $0.00035$ \\
                  & $\Omega$ & $0.01252$ & $0.01053$ & $0.01500$ & $0.00689$ \\
 \hline\hline
\end{tabular}
\end{center}
\vspace{7mm}

It is immediately apparent that, at any fixed time, and for most
categories of error
analysis, the error in $\psi$ is smallest,
and the error in $\omega$ is largest. In keeping with our previous
discussion of the comparison of the spectra, this is associated with the fact
that the fractional error in the higher wavenumber excitations are greater
than that of the lower wavenumbers.

It is also apparent that the errors in the kurtosis are much smaller than
the total rms errors, for a given field. In addition, the error in the
bulk energy
is less than the error in the kurtosis of the velocity.
In fact, the kurtosis errors remain small compared to the total rms error,
especially for the vorticity. The reasons for this appear to be that
the structures, and the distribution of structures in the LBE run
remain quite close to their spectral method counterparts.
However, the exact positions of the structures become different in the LBE
case, relative to the spectral case.
This disparity appears first in the high wavenumber structures, and later on
in the large scale structures, so that by $t=80$ (see Fig. 3d) the large
vortices
that remain are {\it not} at the same locations in the two runs.
Nevertheless the spectra remain very close (see Fig. 2).
As with the spectra, the kurtosis calculation is not sensitive to the
position of structures, but only to their magnitude and shape, and, in a
statistical sense, to the distribution of shapes.
Evidently, the distribution of excitations, in both wavenumber and real space,
remains relatively close for the two methods. The largest error appears
to be in the {\it position} of the vorticity structures, and the large
increase in the error at later times is associated with the progressive
drift in position of the LBE relative to the spectral results.

The origin of the drift in vortex positions, while bulk quantities,
shapes and spectra remain
fairly accurate, can be attributed to several possible causes. First of all,
to compare the methods, we needed to reconcile the LBE timescale with the
spectral (fluid) timescale. This was accomplished (see Section 3) in the
present study by computing
a conversion factor at $t=0$ giving the ratio of the characteristic
time units, involving the rms fluid velocity fluctuation.
The latter quantity changes in time, but this change would not produce
a difference in the results of the two methods if the fluid kinetic
energy and the enstrophy
remained exactly equal for the two cases.  However, there is a small
difference in the energies and enstrophies (see Fig. 1 and Table I), and
this causes a slight inaccuracy in the times at which we compared the results.
As these ``clocks'' drift apart, so do the positions of the
vortices at the times at which we compare them.
This part of the positional drift may be operational in our study, rather
than intrinsic to the differences in the numerical methods, and could,
in principle, be reduced by a more sophisticated, and more difficult,
analysis of the data. A second cause of the positional drift, is closely
related to the first, but is of physical origin. Specifically, we have
argued in Sec. 6, that some of the small departures from incompressible flow
in the LBE method may be a real physical effect, that of nearly
incompressible flow, which the LBE represents reasonably well, but which is
absent in the purely incompressible spectral run. The effects of the small
amount of compressible flow may include differences in the decay of energy
in the two cases, as well as differences in the position of vortices, even
at the same physical time into each run. In this perspective, the
positional drift, as well as other differences in the results of the
two methods, may be attributable to compressive effects, and not numerical
error. We note that both of these possible sources of differences in
the methods, are expected to have greater influence on the total rms error
than on the bulk errors or the kurtosis error.
This is because each of them induce small changes in the effective times of a
comparison. During this small time increment
the position of vortices vary rapidly compared to changes in the spectra,
or compared to changes in their shapes (except possibly at times of vortex
collisions). The total rms error is extremely sensitive to exact positions
of all structures in the simulation domain, even if the structures
are otherwise accurately represented.

We are led to the conclusion that the LBE scheme
has matured to the point that it offers an alternative method for solving
incompressible flow problems with reasonably high accuracy.
In particular, the above error analysis suggests that the LBE approach gives
relatively good results for bulk
quantities such as energy, for wavenumber spectra
and for measures of distributions such as kurtosis.
Although contour plots show great similarity in spectral and LBE cases, there
is, evidently, a growing drift in relative positions
of vortex structures in the
 two cases.
However, for turbulence calculations, the importance of exact positions
of the vortices is rarely considered central, while spectra, energy decay
rates,
and statistics such as kurtosis are of great interest.
Moreover, we find some indication that the scheme also offers
quantitative information
concerning the small effects of compressibility, including ``pseudosound''
density fluctuations associated with the incompressible flow, and accompanying
acoustic waves.
As far as efficiency is concerned, we note that,
for these resolutions and at the Mach number used, the $256^2$ spectral
run ``costs'' about $6$ cpu minutes per characteristic time,
whereas the $512^2$ LBE run ``costs''
about $8$ min per characteristic time on the San Diego Cray YMP.
Thus, the LBE is of comparable efficiency, and may fare better than the
incompressible spectral code in a parallel implementation. However,
one should also note that the timings of a {\it compressible} spectral
code would be expected to be about a factor of $M^{-1}$ longer to resolve
acoustic frequencies. Consequently, if the small compressible effects are
required, the LBE may already be more efficient.

The particular LBE model we have used is the product of several
refinements to the method. These include corrections to the pressure
that enforce a particular (isothermal) equation of state, and the use of
a single time relaxation procedure for handling
the collisional approach to local equilibrium. Further refinements and
extensions are also feasible as well. In particular, the pressure
can, in principle, be further modified to include an independent temperature
variable, so that a full ideal gas equation of state can be implemented.
In addition, the method can be modified \cite{accd}
to allow for higher Mach number flows, and even transonic flows, to be
computed.
However, this has not been attempted here, in view of our goal of
comparison with an incompressible solution, approached through a low Mach
number flow.

\section*{Acknowledgments}
We are grateful to Dr. Hudong Chen for a number of useful discussions.
This work was supported at Bartol by NSF Grant ATM-8913627 and by the
NASA through Innovative Research Program Grant NAG-W-1648 and Space
Physics Theory Program Grant NAG-5-1753, and at Dartmouth by NASA
grant NAG-W-710 and the U.S. Department of Energy Grant DE-FG02-85ER-53194.
One of us (SC) was partly supported by the Department of Energy through
Los Alamos National Laboratory.
All computations were done on the CRAY-YMP at the San Diego Supercomputer
Center.
\section*{Appendix}
In developing the LBE theory it is of interest to understand the relationship
the theory has to kinetic theory of ordinary gases, in addition to
evaluating the computational method itself. In this respect the LBE method
described herein possesses some properties that are unusual from the
ideal gas kinetic theory perspective. Specifically, the present model is
developed to arrive at a useful computational representation
of incompressible flow, evidenced by the emergence of Eq. (5) at lowest
order in the Chapman Enskog expansion, and also at leading order in
a Mach number expansion. However, particularly in view of recent efforts
\cite{sycnew} to employ related LBE methods to flows that may be
strongly compressible, it is important to examine features
of the method, such as the viscosity, when compressible effects are included.
Although a complete examination of these effects has yet to be completed,
we have noted the following disparity between the simple STRA LBE method and
ordinary gas kinetic theory.

In the kinetic theory of simple gases, the kinematic viscosity is expected
to be dependent upon density, approximately as $\nu = \mu/\rho$, where the
molecular viscosity $\mu$ is approximately independent of
density\cite{Batch,Huang}.
This scaling emerges because one finds that $\mu \propto \rho v_{th} \lambda$,
where $v_{th}$ is a thermal speed (roughly analogous to the LBE lattice
streaming speed) and $\lambda$ is a collisional mean free path, related
to a collision time $\tau_c$ by
$\lambda = v_{th} \tau_c$. In spite of what appears
as an explicit
linear dependence of $\mu$ upon $\rho$, it is a familiar result
that the molecular
viscosity is nearly density independent because $\lambda$
(or, equivalently $\tau_c$) scales as $\propto 1/\rho$.
More precisely, on the basis of
kinetic theory, molecular viscosity is independent of density for a
fixed temperature, a fact originally noted by Maxwell, and born out
in standard kinetic theory calculations (e.g.,\cite{Huang}).
However, when such calculations are carried out with a single time
relaxation approximation to the collision operator (with relaxation time
$\tau_c$), the correct scaling is obtained only by associating with
$\tau_c$ an inverse proportionality with
density.

The STRA LBE method used here and elsewhere\cite{ccmm,qian,ccm,sycnew}
differs from the ordinary gas kinetic theory result in that the relaxation
time has typically been chosen as a density independent constant. Consequently,
there are features of the LBE viscosity that differ from the ordinary gas
situation. Most importantly, the
molecular viscosity $\mu$ is {\it not} independent of density, essentially
because the combination $\rho\tau$ still depends on density.
The molecular viscosity cannot be
immediately ``pulled through'' spatial derivatives, divided by $\rho$, and
renamed as the kinematic viscosity $\nu$. Instead there
are also new terms that appear, all of which involve $\nabla \rho$. This
changes the form of the compressible dissipation terms (${\bf D}$ in Eq. 14)
to something other than the precise form expected for a compressible ideal
gas. However, these additional terms, according to the nearly incompressible
flow theory reviewed in Sec. 6, involve two more factors of
Mach number than do the ``usual'' terms in the viscosity. Thus, the added
effects do not directly or significantly influence the incompressible flow
component of the LBE in the
nearly incompressible regime.

These differences reflect
the fact that in LBE theory, in contrast to ordinary gases
(as well as cellular automata \cite{fhp1,fhp2,wolf}), the collisional mean free
part is not determined by actual collisions that occur in the dynamics. Instead
the ``collision rate'' is determined by the selected relaxation
parameter that controls the rate of approach to local equilibrium. This
parameter $\tau$ is
externally controlled and is arbitrary within the bounds set by
stability conditions for the LBE dynamical equation.
Accordingly, the constant STRA collision operator
is adequate, and perhaps
also an efficient way, to compute incompressible or nearly incompressible
flow with an LBE scheme. However, an improvement may be desirable for
LBE schemes that are designed for higher Mach number flows that admit more
effects of compressibility\cite{sycnew}.
In particular, the STRA model can be modified by choosing
$\tau = \tau_0 \rho_0/\rho$ with $\tau_0$ a constant time scale, $\rho_0$ the
mean density, and $\rho$ the local value of density. This modification
is expected to bring a compressible LBE scheme into closer agreement
with the kinetic physics of an ideal gas, particularly with regard to the
structure and value of the viscous transport coefficients.




\section{Figure Captions}
\begin{description}

\item Fig. 1 Time history of a) energy, b) enstrophy, c) palinstrophy,
and d) the next higher order moment, q-enstrophy ($\sim k^4 \omega(k)^2$).
Continuous line corresponds
to the LBE simulation. Departures are noticeable for the higher moments only.

\item Fig. 2 Wavenumber energy spectra for times a) $0$, b) $5$, c) $49$ and
d) $80$, for both spectral and LBE simulations. Continuous line
corresponds to the LBE simulation. The spectra for $t=0$ are identical for
both runs by construction.

\item Fig. 3 Isovorticity contour plots for times a) $1$, b) $5$,
c) $17$, and d) $80$.
Dashed lines correspond to negative values of vorticity.
The values for the contours are the same for all cases. Strikingly similar
features can be found for the LBE simulation as compared with the spectral
simulation.

\item Fig. 4 Correlation between $\omega$ and $\psi$, and between
$\omega$ and $\sinh(|\beta|\psi)$ as a function of time. Continuous
line corresponds to the LBE simulation.

\item Fig. 5. Near incompressibility of the LBE run.
a) time history of the rms transverse velocity normalized
by the sound speed $U_\perp/C_s$. This quantity remains approximately
constant, and equal to the initial Mach number $M=0.069$.
b) $U_L/U_\perp$ as a function of time, where $U_L$ is the rms longitudinal
 velocity. This ratio is clearly bounded by $M$, as required
for approaching incompressibility.
c) Density fluctuations divided by $\rho_0$ as a function of time for the
LBE simulation. $\rho$, $\rho^\infty$ and $\rho\prime$ correspond to
the total density, the ``incompressible'' density, and density
fluctuations associated with acoustic waves, respectively (see text). All
 fluctuations are $O(M^2)$ ($M^2 = 0.0048$), consistent with nearly
incompressible theory.
\end{description}

\section{Table Captions}
\begin{description}

\item Table. 1. Normalized differences between the spectral run and
the LBE run for various quantities for $t=1$, $10$, $50$ and $100$.
$\varepsilon$ is the total rms error, whereas
$\Delta \Phi/\Phi=|\Phi_{SP}-\Phi_{LBE}|/\Phi_{SP}$. Large differences in
$\varepsilon$ are due mainly to a drift in vortex positions. Differences
are significantly reduced for the two lower sections of the table that
show errors in quantities that are independent of the exact distribution
of vorticity but are, instead, sensitive to the shape of the vortices.

\end{description}


\thebibliography{99}

\bibitem{fhp1} U. Frisch, B. Hasslacher and Y. Pomeau,
``Lattice-Gas Automata for the Navier-Stokes Equation'',
Phys. Rev. Lett. {\bf 56}, 1505 (1986).

\bibitem{fhp2}U. Frisch, D. d'Humi\`{e}res, B. Hasslacher,
P. Lallemand, Y. Pomeau, J.-P. Rivet,
``Lattice Gas Hydrodynamics in Two and Three Dimensions'',
Complex Systems, {\bf 1}, 649-707 (1987).

\bibitem{wolf}S. Wolfram,
``Cellular Automaton Fluids 1: Basic Theory'',
J. Stat. Phy. {\bf 45}, 19-74 (1986).

\bibitem{hudong-bill}H. Chen and  W. H. Matthaeus,
``New Cellular Automaton for Magnetohydrodynamics'',
Phy. Rev. Lett., {\bf 58}, 1845(1987). H. Chen, W. H. Matthaeus and
L. W. Klein,
``An Analytic Theory and Formulation of a Local Magnetohydrodynamics
Lattice Gas Model'',
Phys. Fluids, {\bf 31}, 1439-1445 (1988).

\bibitem{roth1}D. H. Rothman and J. M. Keller,
``Immiscible Cellular-Automaton Fluids'',
J. Stat. Phys. {\bf 52}, 1119 (1988).
J.A. Somers and P.C. Rem,
``Analysis of Surface Tensions in Two-Phase Lattice Gases'',
Physica D,{\bf 47}, 39 (1991).
S. Chen, G. D. Doolen,  K. Eggert, D. Grunau and E. Y. Loh,
``A Local Lattice Gas Model for Immiscible Fluids'',
Phys. Rev A, {\bf 43}, 245 (1991).



\bibitem{roth2}D. H. Rothman,
``Cellular-Automaton fluids: a Model for Flow in Porous Media'',
Geophysics {\bf 53}, 509 (1988).

\bibitem{schen2}S. Chen, K. Diemer, G. D. Doolen, K. Eggert, and
S. Gutman and B. J. Travis,
``Lattice Gas Automata for Flow through Porous Media'',
Physica D, {\bf 47} 72  (1991).

\bibitem{macn-zan}G. McNamara and G. Zanetti,
``Use of the Boltzmann Equation to Simulate Lattice-Gas Automata'',
Phys. Rev. Lett.,
 {\bf 61}, 2332 (1988).

\bibitem{hig-jim}F. Higuera and J. Jimenez,
``Boltzmann Approach to Lattice Gas Simulations'',
Europhys. Lett.,
{\bf 9}, 663 (1989).

\bibitem{ccmm} S.Chen, H.Chen, D.Mart\'{\i}nez and
W.H.Matthaeus,
``Lattice Boltzmann Model for Simulation of Magnetohydrodynamics'',
Phys. Rev. Lett {\bf 67},3776 (1991).

\bibitem{qian}Y.H.Qian, D.d'Humieres and P.Lallemand,
``Lattice BGK Models for Navier-Stokes Equation'',
Europhys. Lett.,
{\bf 17}, 479 (1992).

\bibitem{ccm}H.Chen, S.Chen and W.H.Matthaeus,
``Recovery of the Navier-Stokes Equations using a Lattice-Gas Boltzmann
Method', Phys. Rev A, {\bf 45}, 5339 (1992).

\bibitem{schen4}S.Chen, Z.Wang, X.Shan and G.Doolen,
``Lattice Boltzmann Computational Fluid Dynamics in Three-Dimensions'',
J. Stat. Phys. {\bf 68},
379 (1992).

\bibitem{BGK} P. Bhatnagar, E.P. Gross and M.K. Krook,
``A Model for Collision Processes in Gases. I. Small Amplitude Processes
in Charged and Neutral One-Component Systems'',
Phys. Rev.,
{\bf 94} 511 (1954).

\bibitem{DahlMontDool} J.P. Dahlburg, D. Montgomery and G. Doolen,
``Noise and Compressibility in Lattice-Gas Fluids'',
Phys. Rev. A {\bf 36}, 2471 (1987).

\bibitem{schen5}S. Chen, Zhen-su She, L. C. Harrison and
G. D. Doolen,
``Optimal Initial Conditions for Lattice-Gas Hydrodynamics'',
Phys. Rev. A, {\bf 39}
2725 - 2727, (1989).

\bibitem{accd} F.J. Alexander, H. Chen, S. Chen and G.D. Doolen,
Phys. Rev A,
``Lattice Boltzmann Model for Compressible Fluids'',
{\bf 46}, 1967 (1992).

\bibitem{koelman} J.M.V.A. Koelman,
``A Simple Lattice Boltzmann Scheme for Navier-Stokes Fluid Flow'',
Europhys. Lett.,
{\bf 15}, 603 (1991).

\bibitem{ting} A.C. Ting, W.H. Matthaeus and D. Montgomery,
``Turbulent Relaxation Processes in Magnetohydrodynamics'',
Phys. Fluids {\bf 29}, 3261 (1986).

\bibitem{KraichnanandMont} R.H. Kraichnan and D. Montgomery,
``Two-Dimensional Turbulence'',
Rep. Prog. Phys. {\bf 43}, 547 (1987).

\bibitem{bre-hai} F.D. Bretherton and D. Haidvogel,
``Two-Dimensional Turbulence Above Topography'',
J. Fluid Mech. 78, 129 (1976).

\bibitem{mon-tur-va} D. Montgomery, L. Turner, and G. Vahala,
``Three-Dimensional Magnetohydrodynamics Turbulence in Cylindrical Geometry'',
Phys. Fluids 21, 757 (1978).

\bibitem{whm-montg}  W.H. Matthaeus and D. Montgomery,
``Selective Decay Hypothesis at High Mechanical and Magnetic Reynolds
Numbers'',
Ann. N.Y. Acad. Sci. 357, 203 (1980).

\bibitem{us1} W. H. Matthaeus, W. T. Stribling, D. Mart\'{\i}nez, S. Oughton,
and D. Montgomery,
``Selective Decay and Coherent Vortices in Two-Dimensional Incompressible
Turbulence'', Phys. Rev. Lett. 66, 2731 (1991).

\bibitem{us2} W.H.Matthaeus, W.T.Stribling, D. Mart\'{\i}nez, S.Oughton,
and D.Montgomery,
``Decaying, Two-Dimensional, Navier-Stokes Turbulence at Very Long Times'',
Physica D 51, 531 (1991).

\bibitem{us3} D.Montgomery, W.H.Matthaeus, W. T. Stribling, D. Mart\'{\i}nez,
and S. Oughton,
``Relaxation in Two-Dimensions and the ``Sinh-Poisson'' Equation'',
Phys. Fluids A4, 3 (1992).

\bibitem{joy-mon}  G.Joyce and D.Montgomery,
``Negative Temperature States for the Two-Dimensional Guiding-Center Plasma'',
J. Plasma Phys. 10, 107 (1973).

\bibitem{mon-joy} D.Montgomery and G. Joyce,
``Statistical Mechanics of ``Negative Temperature'' states'',
Phys. Fluids 17, 1139 (1974).

\bibitem{smith}  R.A. Smith,
``Maximization of Vortex Entropy as an Organizing Principle in Intermittent,
Decaying, Two-Dimensional Turbulence'', Phys. Rev. A43, 1126 (1991).

\bibitem{mon-tur-va2} D.Montgomery, L.Turner, and G.Vahala,
``Most Probable States in Magnetohydrodynamics'',
J.Plasma Phys. 21, 239 (1979).

\bibitem{ambr-va} J.Ambrosiano and G. Vahala,
``Most Probable Magnetohydrodynamic Tokamak and Reversed Field Pinch
Equilibria'', Phys. Fluids 24, 2253 (1981).

\bibitem{mon-shan-whm} D. Montgomery, X. Shan, and W.H. Matthaeus,
``Navier-Stokes Relaxation to
Sinh-Poisson States at Finite Reynolds Numbers'',  to be published, 1993.

\bibitem{klainerman-majda}S.Klainerman and A.Majda,
``Compressible and Incompressible Fluids'',
Commun. Pure Appl. Math. {\bf 35}, 629 (1982).
A Majda, {\it Compressible Fluid Flow and Systems of Conservation Laws
in Several Space Variables}, Springer-Verlag, New York, 1984.

\bibitem{gosh-whm}D. S.Ghosh and W.H.Matthaeus,
``Low Mach Number Two-Dimensional Hydrodynamic Turbulence: Energy Budgets
and Density Fluctuations in a Polytropic Fluid'',
Phys. Fluids. A
 {\bf 4}, 148 (1992).

\bibitem{sycnew} F. Alexander, S. Chen and J. Sterling,
``Lattice Boltzmann Thermohydrodynamics'',
Phys. Rev. E (1993),
in press.

\bibitem{Batch} G. K. Batchelor, {\it An Introduction to Fluid Dynamics},
Cambridge Univ. Press, New York, 1988.

\bibitem{Huang} K. Huang, {\it Statistical Mechanics}, John Wiley and Sons,
New York, 1987.
\end{document}